\documentclass[preprint,showpacs]{revtex4}
\usepackage{graphicx}
\usepackage{amsmath}
\usepackage{amssymb}
\usepackage{bm}
\usepackage{color}
\def\l{\langle}
\def\r{\rangle}

\begin{document}
\title{Difference of energy density of states in the Wang-Landau algorithm} 
\author{Yukihiro Komura}
\email{y-komura@phys.se.tmu.ac.jp}
\author{Yutaka Okabe}
\email{okabe@phys.se.tmu.ac.jp}
\affiliation{Department of Physics,
 Tokyo Metropolitan University, Hachioji, Tokyo 192-0397, Japan }

\date{\today}

\begin{abstract}
Paying attention to the difference of density of states, 
$\Delta \ln g(E) \equiv \ln g(E+\Delta E) - \ln g(E)$, 
we study the convergence of the Wang-Landau method.  
We show that this quantity is a good estimator to discuss 
the errors of convergence, and refer to the $1/t$ algorithm. 
We also examine the behavior of the 1st-order transition 
with this difference of density of states 
in connection with Maxwell's equal area rule. 
A general procedure to judge the order of transition is given. 
\end{abstract}

\pacs{05.50.+q 75.40.Mg 05.10.Ln 64.60.De}

\maketitle

The Monte Carlo simulation has become a standard method 
to study many-body problems in physics.  However, we sometimes 
suffer from the problem of slow dynamics in the original 
Metropolis algorithm \cite{metro53}. 
One attempt to conquer the problem of slow dynamics is 
the extended ensemble method; 
one uses an ensemble different from the ordinary 
canonical ensemble with a fixed temperature. 
The multicanonical method \cite{berg91,berg92}, 
the parallel tempering, or the exchange Monte Carlo method 
\cite{hukushima,marinari} and 
the Wang-Landau (WL) algorithm \cite{wl01} 
are examples.  
The WL method is an efficient algorithm 
to calculate the energy density of states (DOS), $g(E)$, 
with high accuracy, and was successfully applied to 
many problems \cite{yamaguchi01,okabe06}.  
The refinement and convergence of the WL method 
were argued \cite{zhou05,lee06}, 
but the convergence property is still a topic 
of discussions \cite{belardinelli07b}.  
The search for optimal modification factor 
was discussed \cite{zhou08}, and in connection with 
the WL method, the $1/t$ algorithm 
\cite{belardinelli07,belardinelli08} was proposed. 
Moreover, \textit{tomographic} entropic sampling scheme has been 
proposed as an algorithm to calculate DOS \cite{dickman}.

In this paper, we investigate the convergence properties of 
the WL method, paying special attention 
to the difference of DOS.  We argue its relevance 
to the 1st-order transition.  We provide a general strategy 
to judge the order of transition.  

Let us briefly review the WL algorithm.  
A random walk in energy space is performed with a probability 
proportional to the reciprocal of the DOS, $1/g(E)$, 
which results in a flat histogram of energy distribution.  
Actually, we make a move based on the transition 
probability from energy level $E_1$ to $E_2$
\begin{align}
  p(E_1 \to E_2) = \min \Big[ 1,\frac{g(E_1)}{g(E_2)} \Big]. 
\end{align}
Since the exact form of $g(E)$ is not known \textit{a priori}, 
we determine $g(E)$ iteratively.  Introducing the modification 
factor $f_i$, $g(E)$ is modified by
\begin{align}
  \ln g(E) \to \ln g(E) + \ln f_i
\end{align}
every time the state is visited. At the same time the energy 
histogram $h(E)$ is updated as
\begin{align}
  h(E) \to h(E) + 1.
\end{align}
The modification factor $f_i$ is gradually reduced to unity 
by checking the `flatness' of the energy histogram. 
The 'flatness' is checked such that the histogram 
for all possible $E$ is not less than some value of 
the average histogram, say, 80\%.
Then, $f_i$ is modified as 
\begin{align}
  \ln f_{i+1} = \frac{1}{2} \ln f_i
\end{align}
and the histogram $h(E)$ is reset.  As an initial value of 
$f_i$, we choose $f_0 = e$; as a final value, we choose 
$\ln f_i = 2^{-26}$, that is, $f_{26} \simeq 1.000 \, 000 \, 01$, for example.

We first treat the Ising model, whose Hamiltonian is given by
\begin{equation}
 \mathcal{H} = -J\sum_{\l i,j \r} \sigma_{i}\sigma_{j}.
 \label{Ising}
\end{equation}
Here, $J$ is the coupling and $\sigma_{i}$ is the Ising spin ($\pm 1$) 
on the lattice site $i$. 
The summation is taken over the nearest neighbor pairs $\l i,j \r$. 
Periodic boundary conditions are employed. 
Throughout this paper, we measure the energy in units of $J$ 
unless specified; in other words, we put $J=1$. 

We calculate $\ln g(E)$ with the use of the WL method, 
and consider the difference of $\ln g(E)$, 
which is defined as 
\begin{align}
 \Delta \ln g(E) \equiv \ln g(E+\Delta E) - \ln g(E).
\label{Delta_gE}
\end{align}
For the Ising model, $\Delta E = 4J$. 
The exact value of $g(E)$ for the two-dimensional (2D) Ising model 
is available due to Beale \cite{beale}.
The deviation of the calculated value of $\Delta \ln g(E)$ 
from the exact value of Beale \cite{beale} 
can be used as a measure of the accuracy of the calculation.

\begin{figure}
\begin{center}
\includegraphics[width=0.6\linewidth]{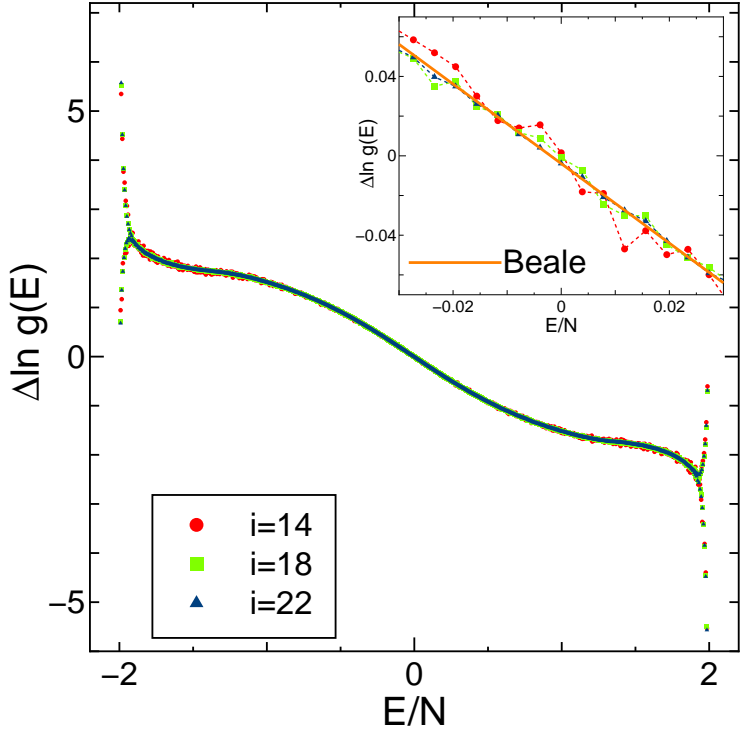}
\end{center}
\caption{(Color online) Plot of $\Delta \ln g(E)$ for the 2D Ising model with $L = 32$. 
Data for $i$ = 14, 18 and 22 are given. In the inset, the enlarged plot 
near $E$ = 0 is shown.  The exact value due to Beale \cite{beale} 
is also given in the inset for comparison.}
\label{fig_1}
\end{figure}

We plot the overall behavior of $\Delta \ln g(E)$ for the 2D 
Ising model with system size $L$ = 32 in Fig.~\ref{fig_1}. 
The data for the modification step $i$ = 14, 18 and 22 
are given for a single measurement.  In the accuracy of this plot, 
little difference in $i$ is appreciable except for small and large $E$. 
The enlarged plot near $E = 0$ is given in the inset of Fig.~\ref{fig_1}, 
and the data for $i$ = 14, 18 and 22 are 
compared to the exact value of Beale \cite{beale}.  
We see that the calculated value of $\Delta \ln g(E)$ approaches 
the exact value as the modification factor $f_i$ approaches 1.
The deviation becomes smaller as $i$ increases. 
The advantage of using Eq.~(\ref{Delta_gE}) is that 
we can directly discuss the error of DOS without 
caring about the normalization of $g(E)$. 
Since the transition probability depends on 
the difference of $\ln g(E_1)$ and $\ln g(E_2)$, 
this quantity of difference is essential in the method 
calculating the energy DOS compared to $g(E)$ itself. 
We note that the quantity of difference was also used 
in the argument of accuracy and convergence of 
the WL method by Morozov and Lin \cite{morozov}.

To see the convergence of errors more explicitly, we consider 
the total sum of the squared error of 
$\Delta \ln g(E)- \Delta \ln g(E)_{\rm exact}$; 
\begin{align}
  \Delta^2 \equiv \frac{1}{N-4} \sum_{E = -2JN+8J}^{2JN-12J} 
         \big( \Delta \ln g(E)- \Delta \ln g(E)_{\rm exact} \big)^2
\label{error}
\end{align}
For the 2D Ising model, we note that 
$\Delta \ln g(-2JN+8J)_{\rm exact} = - \Delta \ln g(2JN-12J)_{\rm exact} 
= \ln 2$.

\begin{figure}
\begin{center}
\includegraphics[width=0.6\linewidth]{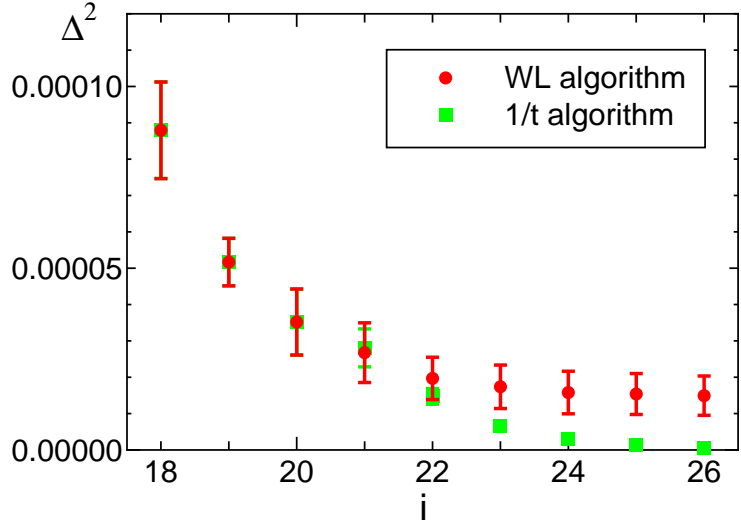}
\end{center}
\caption{(Color online) Convergence of errors, $\Delta^2$, for the 2D Ising model 
with $L = 32$ as a function of the modification step $i$.  
The convergence of the original WL algorithm is compared with 
that of $1/t$ algorithm.  In the $1/t$ algorithm after the rule of 
modification is changed, the meaning of $i$ is such that MCS is $2^i$. 
}
\label{fig_2}
\end{figure}

In Fig.~\ref{fig_2}, we plot $\Delta^2$, Eq.~(\ref{error}), 
as a function of the modification step $i$ up to 26 for $L$ = 32. 
The average is taken for 10 samples. 
We see that $\Delta^2$ becomes smaller with the increase of $i$. 
However, the errors are saturated even though we repeat 
the iteration process up to $i$ = 26. 
Such saturation of convergence of the WL method was pointed out 
by Yan and de Pablo \cite{yan03}. 
To overcome this difficulty, a modified version of the WL algorithm 
in which the refinement parameter is scaled down as $1/t$ 
(with $t$ the Monte Carlo time) was proposed 
\cite{belardinelli07,belardinelli08}. 
It is interesting to compare the performance of the $1/t$ algorithm 
and that of the original WL method in this quantity of difference of DOS. 
In the $1/t$ algorithm, starting from the same condition as 
the original WL algorithm, the modification factor $\ln f_i$ is 
reduced as $1/t$ instead of checking the flatness condition 
after the condition $\ln f_i \le 1/t$ is satisfied.  
The final value of $\ln f$ should be fixed from the outset. 
In Fig.~\ref{fig_2}, we also plot the data for the $1/t$ algorithm. 
In the case of $L=32$, the modification process is changed 
from the original WL scheme to the $1/t$ one around $i=$21 or 22. 
In the range of $1/t$ scheme the actual MCS is fixed as $2^i$, 
which is different from the case of the original WL scheme. 
We clearly confirm the efficiency of the $1/t$ algorithm. 
In the discussion of the convergence of $1/t$ algorithm, 
the quantity $\ln g(E)-\ln g(E_{\rm ground})$ 
was used \cite{belardinelli07b}.  
The quantity given by Eq.~(\ref{Delta_gE}) is 
more flexible as it can be treated even if the ground state 
of the system is unknown as spinglass problems. 

We can consider the deviation from the exact value, 
as in Eq.~(\ref{error}), for the 2D Ising model. 
In order to investigate the convergence behavior of the system 
whose exact $g(E)$ is not available, we may employ another 
strategy.  For example, we may consider the relative error 
of the data for $i$ and those for $i-1$. 
We leave the detailed analysis to a separate publication. 

Next we deal with the 2D ten-state Potts model, which is a typical 
model to exhibit the 1st-order transition. 
This model was used to show the effectiveness of the multicanonical 
simulation by correctly estimating 
the interfacial free energy \cite{berg92}, which was later proved 
by the explicit formula \cite{borgs}.  
The Hamiltonian of the $q$-state Potts model is given by
\begin{equation}
 \mathcal{H} = J\sum_{\l i,j \r} \big[ 1 - \delta_{S_{i},S_{j}} \big].
 \label{Potts}
\end{equation}
Here, $S_{i}$ is the Potts spin which takes $1, \cdots, q$. 
We note that for $q = 2$ the Potts model becomes the Ising model, 
although the unit of $J$ in Eq.~(\ref{Potts}) for the Potts model 
is twice as $J$ in Eq.~(\ref{Ising}) for the Ising model.

\begin{figure}
\begin{center}
\includegraphics[width=0.55\linewidth]{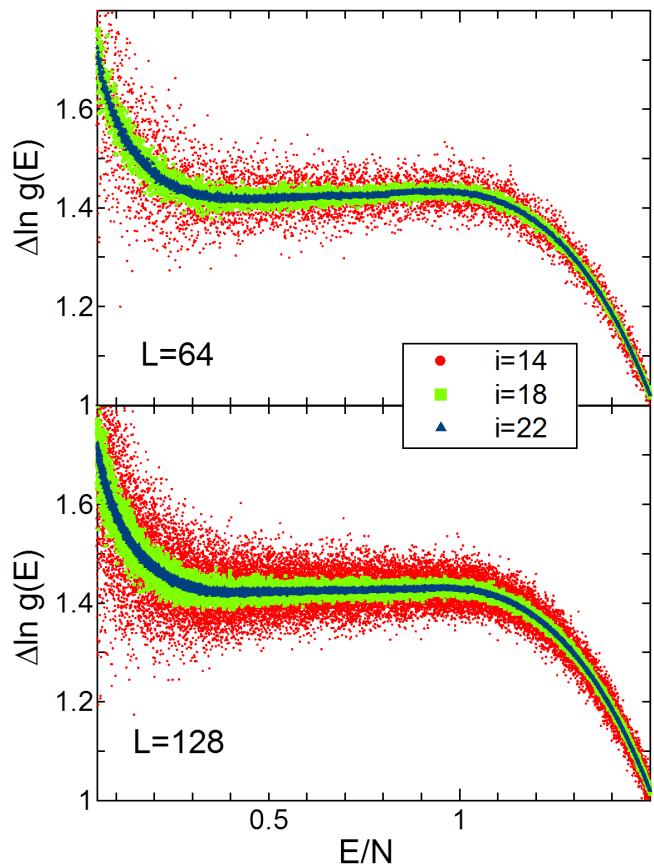}
\end{center}
\caption{(Color online) Plot of $\Delta \ln g(E)$ of the 2D ten-state Potts model 
for $L$ = 64 (upper) and $L$ = 128 (lower) as a function of $E/N$. 
The data for the modification factor $f_i$ 
with $i$ = 14, 18 and 22 are given.
}
\label{fig_3}
\end{figure}

We plot the difference of DOS, Eq.~(\ref{Delta_gE}), of the 2D 
ten-state Potts model in Fig.~\ref{fig_3}. 
The data for $L$ = 64 (upper) and those for $L$ = 128 (lower) are given 
as a function of $E/N$. 
We show how the data converge as $i$ increases by giving 
the data for $i$ = 14, 18 and 22 with a single measurement. 
We clearly see the convergence of errors with the increase of $i$. 

The systems which show the 1st-order transition have double maximum 
structure in the thermodynamic limit at the 1st-order transition 
temperature $T_c$ when we plot the free energy $- \beta F = \ln g(E) - \beta E$ 
as a function of $E$. 
Then, $\Delta \ln g(E)$, which is defined as Eq.~(\ref{Delta_gE}), 
has an $S$-like structure with minimum and maximum.  
We clearly find this structure in Fig.~\ref{fig_3}. 
We note that the overall size dependence is small in this plot, 
but the detailed analysis is given later. 

\begin{figure}
\begin{center}
\includegraphics[width=0.6\linewidth]{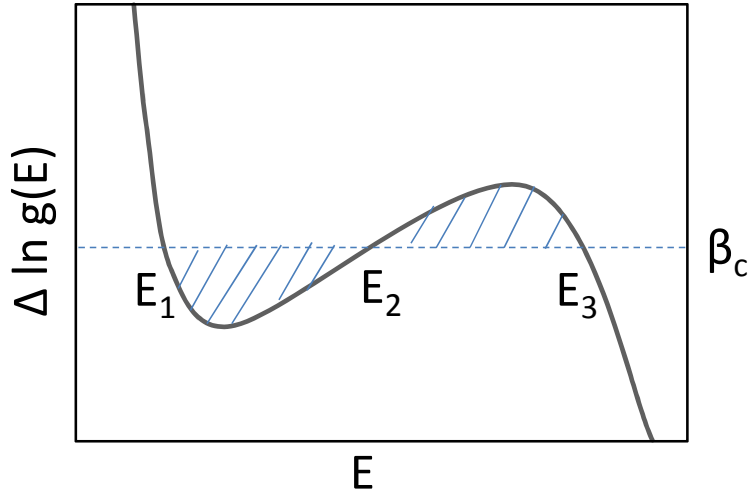}
\end{center}
\caption{(Color online) Schematic illustration of Maxwell's equal area rule.}
\label{fig_4}
\end{figure}

The 1st-order transition temperature, $T_c = 1/\beta_c$, 
can be estimated by Maxwell's rule as in thermodynamics. 
A schematic illustration of Maxwell's rule is shown in Fig.~\ref{fig_4}. 
The value of $\beta$, which separates the shaded region and 
gives the same area, becomes the 1st-order transition temperature 
$\beta_c$.  This equal area rule is proved by the following. 
The condition that the two areas of the shaded region are equal is given by
\begin{align}
 &- \int_{E_1}^{E_2} \frac{d \ln g(E)}{d E} \ dE 
        + \beta (E_2-E_1)  \nonumber \\
 &\quad = \int_{E_2}^{E_3} \frac{d \ln g(E)}{d E} \ dE 
        - \beta (E_3-E_2), 
\label{area}
\end{align}
which leads to the condition that the double maxima in $\ln g(E) - \beta E$ 
take the same value. 
In the thermodynamic limit, the difference $\Delta \ln g(E)$ becomes 
the differential $d \ln g(E)/d E$. 
The area of the shaded region, Eq.~(\ref{area}), 
is related to the interfacial free energy \cite{berg92,borgs}.

\begin{figure}
\begin{center}
\includegraphics[width=0.55\linewidth]{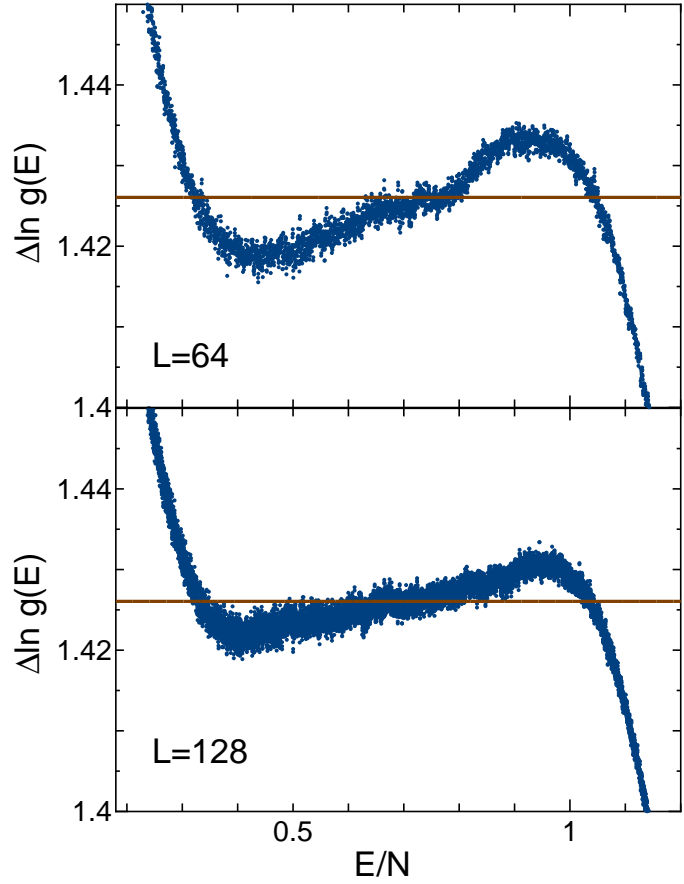}
\end{center}
\caption{(Color online) Enlarged plot of $\Delta \ln g(E)$ of the 2D ten-state Potts model 
for $L$ = 64 (upper) and 128 (lower).  
The modification step $i$ is 22.  The smoothed values 
with moving-average method are given.  
The 1st-order transition temperature 
$\beta_c = \ln(1+\sqrt{10}) = 1.42606$ 
in the thermodynamic limit is also shown by straight line for convenience. 
}
\label{fig_5}
\end{figure}

To see the $S$-like structure explicitly, we make an enlarged plot 
along $y$-axis of $\Delta \ln g(E)$ for $L$ = 64 (upper) and 128 (lower) 
in Fig.~\ref{fig_5}.  The modification step $i$ is 22. 
In this plot we use the data with the smoothing process, 
$(f(E-2\Delta E)+4*f(E-\Delta E)+6*f(E)+
4*f(E+\Delta E)+f(E+2\Delta E))/16$ with $f(E)=\Delta \ln g(E)$, 
to reduce fluctuations. 
For the 2D ten-state Potts model, 
the 1st-order transition temperature is given by 
$\beta_c = \ln(1+\sqrt{10}) = 1.42606$. 
We give this value in Fig.~\ref{fig_5} for convenience; 
we see that Maxwell's rule works. 
We can estimate $\beta_c$ and the interfacial free energy 
from the $S$-like curve for each size. 
We observe the size dependence in Fig.~\ref{fig_5}; 
the area of the shaded region illustrated in Fig.~\ref{fig_4} 
is proportional to $1/L$, 
which reflects on the finite size scaling of the 1st-order 
transition. 

We may provide a general strategy to judge the order of transition 
for any system.  We plot $\Delta \ln g(E)$ and check 
whether there is an $S$-like structure.  
If the system shows the 1st-order transition, 
we can locate the transition temperature by Maxwell's rule. 
The behavior of the 1st-order transition can be observed 
in the early stage of WL iteration, that is, for small $i$.
If we investigate $\ln g(E) - \beta E$ as in usual way, 
we have to search for $\beta$ which gives the same value 
for two maxima. 

To summarize, we have shown that the difference of $\ln g(E)$ 
is a good quantity for the WL method.  
Less attention has been given to the quantity 
$\Delta \ln g(E)$ so far, although some efforts were made 
in the discussion of accuracy and convergence of 
the WL method \cite{morozov}. 
Comparing with the exact value of the 2D Ising model, 
we have shown the convergence property of the WL method.
That is, we have shown how errors become smaller for larger $i$,
where $i$ is the step of the modification factor $f_i$ 
for the criterion of 'flatness' condition. 
We have confirmed the efficiency of the $1/t$ algorithm; 
we have shown that the quantity $\Delta \ln g(E)$ is a good 
estimator for the analysis of errors of the simulation method 
to calculate the energy DOS.

We have also shown that $\Delta \ln g(E)$ is a good 
estimator for the 1st-order transition.  
We have investigated the 2D ten-state Potts model. 
The 1st-order transition is observed in the $S$-like behavior 
of $\Delta \ln g(E)$.  
We have shown that Maxwell's equal area rule determines 
the 1st-order transition temperature.  Although the statement is 
rigorously realized in the thermodynamic limit, 
we observe the behavior of the 1st-order transition 
even for small system size and for small $i$ of the modification step. 
We assert that we provide a general procedure to study 
the order of transition for any system. 

The extension of this calculation to continuous spin models 
is straightforward \cite{zhou06}.  
The application to quantum Monte Carlo simulation 
\cite{troyer03} for checking the order of transition 
is highly desirable. 
The application to first-principle calculation of electric structure 
\cite{eisenbach} and to protein systems \cite{gervais} may be 
other interesting topics.s

Before closing, we mention about the calculation techniques. 
We have used the parallel calculation with multiple random walkers 
for the WL algorithm using the GPU (graphic processing unit) 
with CUDA (common unified device architecture).  
The details of the GPU-based calculation will be given elsewhere.

We thank Tasrief Surungan for valuable discussions. 
This work was supported by a Grant-in-Aid for Scientific Research from
the Japan Society for the Promotion of Science.


\begin{thebibliography}{99}

\bibitem{metro53} 
 N. Metropolis, A.W. Rosenbluth, M.N. Rosenbluth, A.H. Teller, 
 and E. Teller, 
 J. Chem. Phys. {\bf 21}, 1087 (1953).

\bibitem{berg91}  B. A. Berg and T. Neuhaus, 
 Phys. Lett. B {\bf 267}, 249 (1991).
\bibitem{berg92}  B. A. Berg and T. Neuhaus, 
 Phys. Rev. Lett. {\bf 68}, 9 (1992).
\bibitem{hukushima}  K. Hukushima and K. Nemoto, 
 J. Phys. Soc. Jpn. {\bf 65}, 1604 (1996).
\bibitem{marinari} E. Marinari,
 in {\it Advances in Computer Simulation}, 
 edited by J. Kert\'esz and I. Kondor,  
 (Springer-Verlag, Berlin, 1998), p. 50.

\bibitem{wl01}
 F. Wang and D.P. Landau,  
 Phys. Rev. Lett. {\bf 86}, 2050 (2001); 
 Phys. Rev. E {\bf 64}, 056101 (2001).

\bibitem{yamaguchi01}
 C. Yamaguchi and Y. Okabe,
 J. Phys. A: Math. Gen. {\bf 34}, 8781 (2001).
\bibitem{okabe06}
 Y. Okabe and H. Otsuka,
 J. Phys. A: Math. Gen. {\bf 39}, 9093 (2006).

\bibitem{zhou05} 
 C. Zhou and R. N. Bhatt, 
 Phys. Rev. E {\bf 72}, 025701(R) (2005).

\bibitem{lee06} 
 H. K. Lee, Y. Okabe, and D. P. Landau,
 Comp. Phys. Comm. {\bf 175}, 36 (2006).
\bibitem{belardinelli07b} 
 R. E. Belardinelli and V. D. Pereyra, 
 J. Chem. Phys. {\bf 127}, 184105 (2007).
\bibitem{zhou08} 
 C. Zhou and J. Su,
 Phys. Rev. E {\bf 78}, 046705 (2008).

\bibitem{belardinelli07} 
 R. E. Belardinelli and V. D. Pereyra, 
 Phys. Rev. E {\bf 75}, 046701 (2007).
\bibitem{belardinelli08} 
 R. E. Belardinelli, S. Manzi, and V. D. Pereyra, 
 Phys. Rev. E {\bf 78}, 067701 (2008).

\bibitem{dickman}
 R. Dickman and A. G. Cunha-Netto,
 Phys. Rev. E {\bf 84}, 026701 (2011).

\bibitem{beale}
 P. D. Beale, 
 Phys. Rev. Lett. {\bf 76}, 78 (1996).

\bibitem{morozov}
 A. N. Morozov and S. H. Lin,
 Phys. Rev. E {\bf 76}, 026701 (2007);
 J. Chem. Phys. {\bf 130}, 074903 (2009).

\bibitem{yan03}
 Q. Yan and J. J. de Pablo,
 Phys. Rev. Lett. {\bf 90}, 035701 (2003).

\bibitem{borgs}
 C. Borgs and W. Janke, 
 J. Phys. (France) I {\bf 2}, 2011 (1992).

\bibitem{zhou06} 
 C. Zhou, T. C. Schulthess, S. Torbr\"ugge, and D. P. Landau, 
 Phys. Rev. Lett. {\bf 96}, 120201 (2006).
\bibitem{troyer03} 
 M. Troyer, S. Wessel, and F. Alet,
 Phys. Rev. Lett. {\bf 90}, 120201 (2003).
\bibitem{eisenbach}
 M. Eisenbach, C.-G. Zhou, D. M. Nicholson, G. Brown, J. Larkin, 
 and T. C. Schulthess, 
 in: SC, Portland, Oregon, USA, November 14-20, ACM, New York (2009).
\bibitem{gervais}
 C. Gervais, T. W\"ust, D. P. Landau, and Y. Xu,
 J. Chem. Phys. {\bf 130}, 215106 (2009).


\end{thebibliography}
\end{document}